\documentclass[letterpaper, 10 pt, conference]{ieeeconf}  

\IEEEoverridecommandlockouts                              
\usepackage{subcaption}
\usepackage{enumerate}
\usepackage{graphicx}

\title{\LARGE \bf
Which friends are more popular than you? Contact strength and the friendship paradox in social networks}

\author{James P.~Bagrow$^{1}$, Christopher M.~Danforth$^{1}$ and Lewis Mitchell$^{2}$
\thanks{$^{1}$Mathematics \& Statistics, 
Vermont Complex Systems Center,
University of Vermont, Burlington, VT, USA
        {\tt\small \{james.bagrow, chris.danforth\}@uvm.edu}}%
\thanks{$^{2}$School of Mathematical Sciences,
University of Adelaide,
Adelaide 5005,
Australia
        {\tt\small lewis.mitchell@adelaide.edu.au}}%
}

\newcommand{\todo}[1]{}
\renewcommand{\todo}[1]{{\color{red}[[TODO: {#1}]]}}
\newcommand{\LM}[1]{}
\renewcommand{\LM}[1]{{\color{blue}[[LM: {#1}]]}}
\newcommand{\JB}[1]{}
\renewcommand{\JB}[1]{{\color{magenta}[[JB: {#1}]]}}

\newcommand{\kout}[0]{\ensuremath{k_\mathrm{out}}}

\begin{document}

\maketitle
\thispagestyle{empty}
\pagestyle{empty}

\begin{abstract}

The friendship paradox states that in a social network, egos tend to have lower degree than their alters, or, ``your friends have more friends than you do''.
Most research has focused on the friendship paradox and its implications for information transmission, but treating the network as static and unweighted.
Yet, people can dedicate only a finite fraction of their attention budget to each social interaction: a high-degree individual may have less time to dedicate to individual social links, forcing them to modulate the quantities of contact made to their different social ties. 
Here we study the friendship paradox in the context of differing contact volumes between egos and alters, finding a connection between contact volume and the strength of the friendship paradox.
The most frequently contacted alters exhibit a less pronounced friendship paradox compared with the ego, whereas less-frequently contacted alters are more likely to be high degree and give rise to the paradox.
We argue therefore for a more nuanced version of the friendship paradox:
``your closest friends have slightly more friends than you do'',
and in certain networks even:
``your best friend has no more friends than you do''.
We demonstrate that this relationship is robust, holding in both a social media and a mobile phone dataset.
These results have implications for information transfer and influence in social networks, which we explore using a simple dynamical model.

\end{abstract}

\section{INTRODUCTION}

The ``friendship paradox'' is the observation that the friends of individuals tend to have more friends on average than the individuals themselves. 
This notion was first articulated for social networks \cite{feld1991your} but it plays a role on other settings as well~\cite{feld1977variation,hemenway1982your,good1983good}. 
The friendship paradox stands alongside other famous features of social networks: the strength of weak ties, that people tend to gain novel information not from close friends but from weak acquaintances~\cite{granovetter_weakties_1973}, and Dunbar's number, which argues that human's have a cognitive upper limit on the effective size of an individual's social circle~\cite{dunbar1993coevolution}. 
Researchers have studied and validated these properties in a number of modern datasets, including Twitter~\cite{kwak2010twitter,Hodas13icwsm}, Facebook~\cite{ellison2007benefits,ugander2011anatomy} and mobile phones~\cite{onnela2007structure,bagrowDisaster2011pone,gao2014quantifying}.
Network degree relates to many personality traits such as introversion~\cite{pollet2011extraverts}, 
and perceptual biases also factor into the effects of the friendship paradox~\cite{zuckerman2001makes}.
All of these phenomena in concert provide a deeper understanding of how social networks form under cognitive and attentional limits, how information can move between individuals via the network, and to what extent the individual's embedded view corresponds to the actual network or information within it.

The friendship paradox along with the strength of weak ties play significant roles in how one is exposed to information socially.
By increasing the size of his or her social circle, an individual can gain access to a potentially overwhelming array of information.
But in the modern ``attention economy'', limited attention and bombardment of information may dull the effectiveness of social information spread by making it more difficult to exchange information with weak ties~\cite{hodas2012visibility}.
This can be due to reduced opportunities for communication and, 
if those weak ties are popular individuals, limited attention to spend on any one of their social ties. 
Therefore, it is crucial to understand the interplay between an individual and his or her social ties, not simply from the static features of those ties but also from the dynamics and relative frequencies of those interactions.
Focussing as they do on the structural properties of static, unweighted social networks, previous analyses ignore valuable information on the nature and strength of social ties.

In this work, we study two social network datasets, one derived from Twitter and one from mobile phone activity. 
We move beyond the classic friendship paradox analysis to consider what effects the volume of contact between egos and alters may have on the paradox. 
We show that there is a significant relationship between these quantities, in particular that the more heavily or frequently contacted alters tend to have fewer social ties than the less frequently contacted alters. 
This implies that much of the driving force behind the friendship paradox is due to these weak ties. 
Expanding upon this analysis, we investigate the generality of our results by studying similarities and differences between the properties of these two very different datasets and determining how those differences track with the relationship between contact volume and the number of friends of alters.
For example, the outdegree of Twitter users is more heavily right-skewed than mobile phone users, yet both sets of users display the same contact volume--outdegree relationship. 
We apply a null model to show the significance of this relationship while accounting for data sampling effects.
The friendships paradox plays a role in information diffusion over a social network alongside other features such as the strength of weak ties. We use a dynamical model of a simple information diffusion process to show that the relationship plays a significant role in slowing down the spread of information.
The work we present here moves beyond static network properties and strengthens the connection between the friendship paradox, 
the strength of weak ties, 
and Dunbar's number.


\section{DATASETS}

We focused on two social network datasets, one taken from Twitter.com social media users and one from Mobile Phone billing records. We extracted thousands of egocentric networks from each dataset, gathering the outdegrees of egos and alters, and ranking alters by the volume of contact from the ego (rank-1 is the most frequently contacted alter).

\subsection{Twitter}

\begin{itemize}\itemsep=0pt
\item Here the outdegree $\kout$ was the number of followers of users and alters were ranked by the number of at-mentions from the ego.
\end{itemize}
Specifically, we collected the Twitter dataset by selecting a random sample of individuals for study from the 10\% Gardenhose feed collected during the first week of April 2014.
From this, we uniformly sampled individuals who had tweeted in English (as reported by Twitter in the metadata for each tweet) during this time period and, 
to control for confounding effects of the highly right-skewed distribution of Twitter users, 
had 50 $\leq \kout \leq$ 500 followers, 
as reported in the tweet metadata.
We then collected their complete public tweet history
(up to 3200 most recent messages by Twitter's Public REST API limit \cite{TwitterAPI}, excluding retweets. These individuals formed the egos of the dataset. For each ego, we determined through their messages their top-15 most at-mentioned alters, and then proceeded to gather the tweet histories and other metadata (number of followers) of these alters using the Public REST API in the same manner. We limited ourselves to the top-15 alters because gathering the complete tweet history from the API is time-intensive. When finished, this collection process gathered 3,300 individual egos 
and 38,650 alters, 32,980 of which are unique across ego networks.

\textit{Sampling---}
To be retained as part of the Twitter data we required that all egos and alters have posted at least 10k total words in their complete, collected public tweet history.
It is possible that some of the top-15 alters for a given ego do not satisfy this criterion, leading to gaps in that ego's set of alters. However, we still use all at-mentioned alters for ranking purposes, so that if, e.g., an ego with 3 alters has the rank-2 alter excluded from calculations, we still know and retain the correct rank of 3 for the third alter.

\subsection{Mobile Phone}

\begin{itemize}\itemsep=0pt
\item Here $\kout$ was the number of unique recipients of phone calls or text messages and alters were ranked by the number of calls and/or texts they received from the ego.
\end{itemize}
This dataset is comprised of de-identified billing records of approximately 10M subscribers to a mobile phone service provider in a western European country~\cite{gonzalez2008understanding,song2010limits,bagrow2012mesoscopic}. We sampled approximately 90 thousand subscribers from a nine-month period of these billing records to form the egos of the dataset. These records contain the full call histories of each subscriber, from which we can extract three time-series:
(i) timestamps when phone calls were made or text messages sent, to an hourly resolution; (ii) approximate locations from where calls/texts were made, as quantified by the cellular tower transmitting the call/text; and (iii) the recipients of these calls and texts.
To ensure a complete picture of the activities of the egos, we applied the sampling criteria of \cite{song2010limits}: selected egos must place calls from at least two distinct cellular towers during the nine-month data window, and their average call frequency must exceed 0.5 hour$^{-1}$. No constraints were placed on the outdegrees.
For each ego, the third time series was used to reconstruct their corresponding egocentric communication network, up to the 100 most-contacted alters, with call/text recipients forming the ego's set of alters.

\textit{Sampling---}
The mobile phone records are limited to subscribers of a single service provider, which means we only have call histories for subscribers. Yet, subscribers will often call non-subscribers. In these data, non-subscribers are consistently identified for us, but their activities are not captured except when they interact with subscribers. This means that, while all egos in our study have full data, not all alters do, and we cannot compute the outdegree of non-subscriber alters. However, because we rank alters based on the contact volume from the ego, we still rank all alters correctly. To account for this, we only include subscriber alters within our calculations, but using their correct ranks. For example, an ego with three alters where the rank-2 alter is not a subscriber will still accurately provide two data points (ego-alter dyads) for any calculations, the rank-1 alter and the rank-3 alter.

\section{RESULTS}

\begin{figure*}[t!]
    \centering
    \begin{subfigure}[t]{0.5\textwidth}
        \centering
        \includegraphics[width=\textwidth]{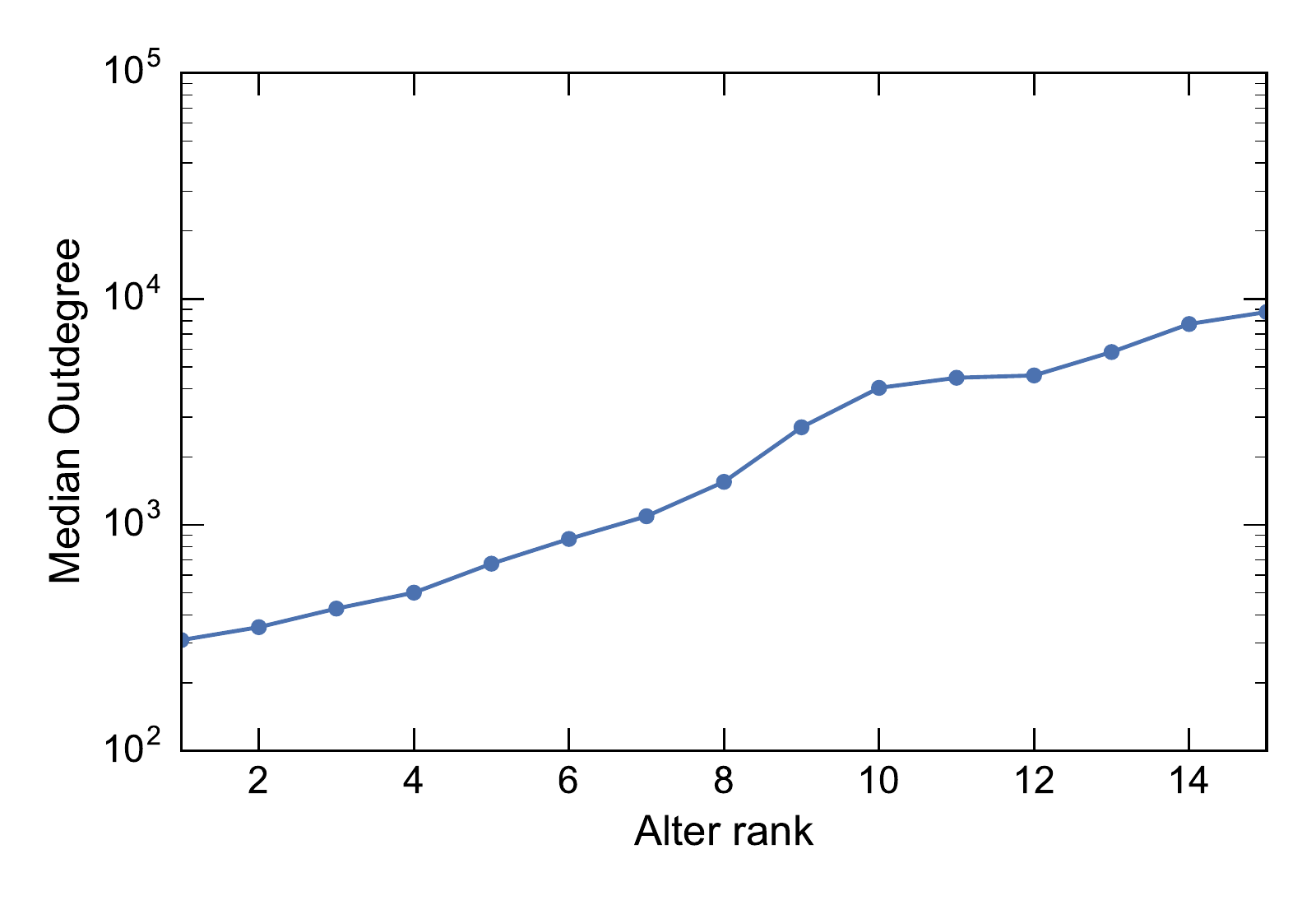}
	    \begin{picture}(0,0)
			\put(-85.5,98){\includegraphics[width=0.42\textwidth,trim=15 35 15 15,clip=true]{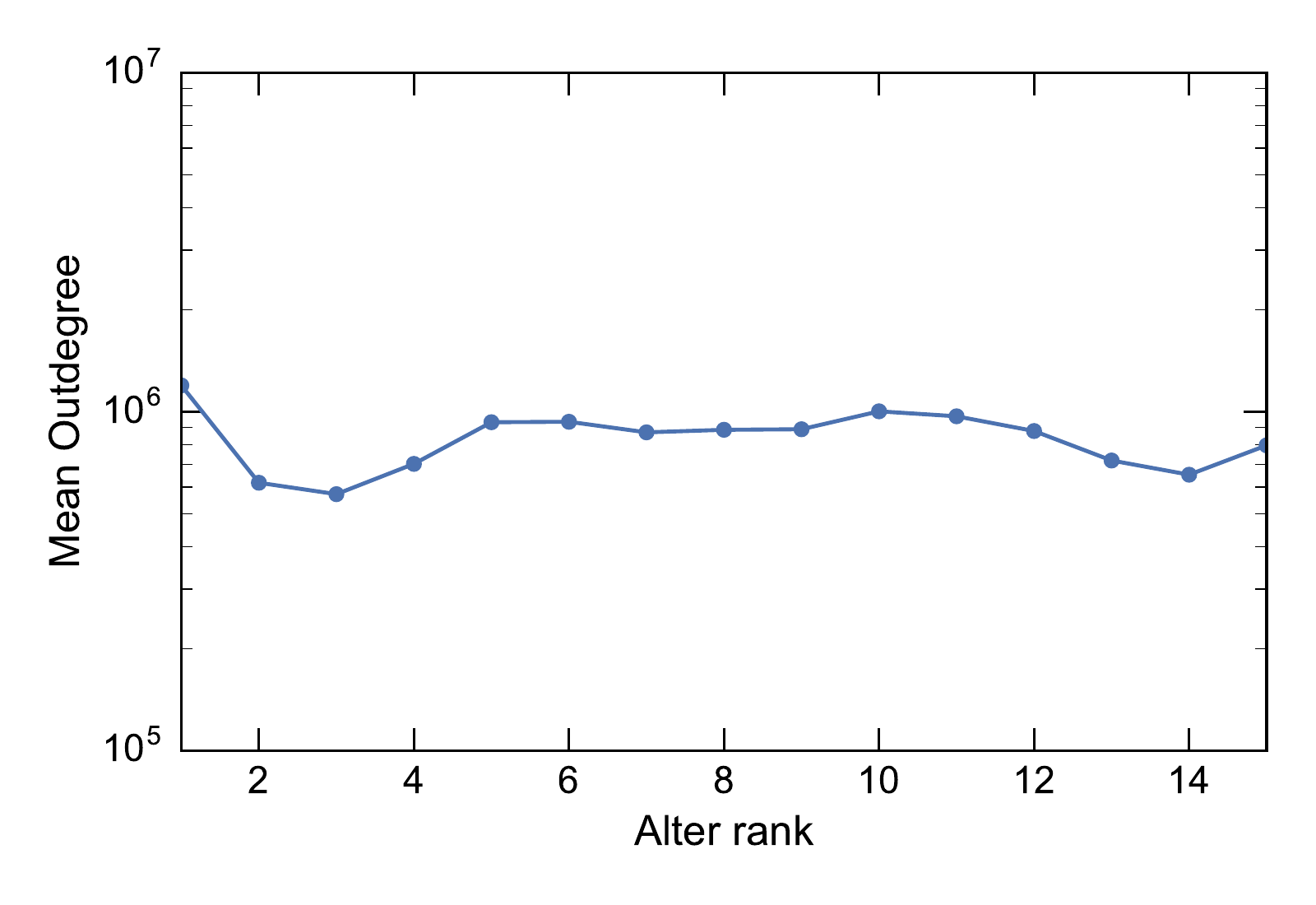}}
		\end{picture}
		\vspace{-1.2em}
        \caption{Twitter\label{subfig:mainResultTwitter}}
    \end{subfigure}%
    ~ 
    \begin{subfigure}[t]{0.5\textwidth}
        \centering
        \includegraphics[width=\textwidth]{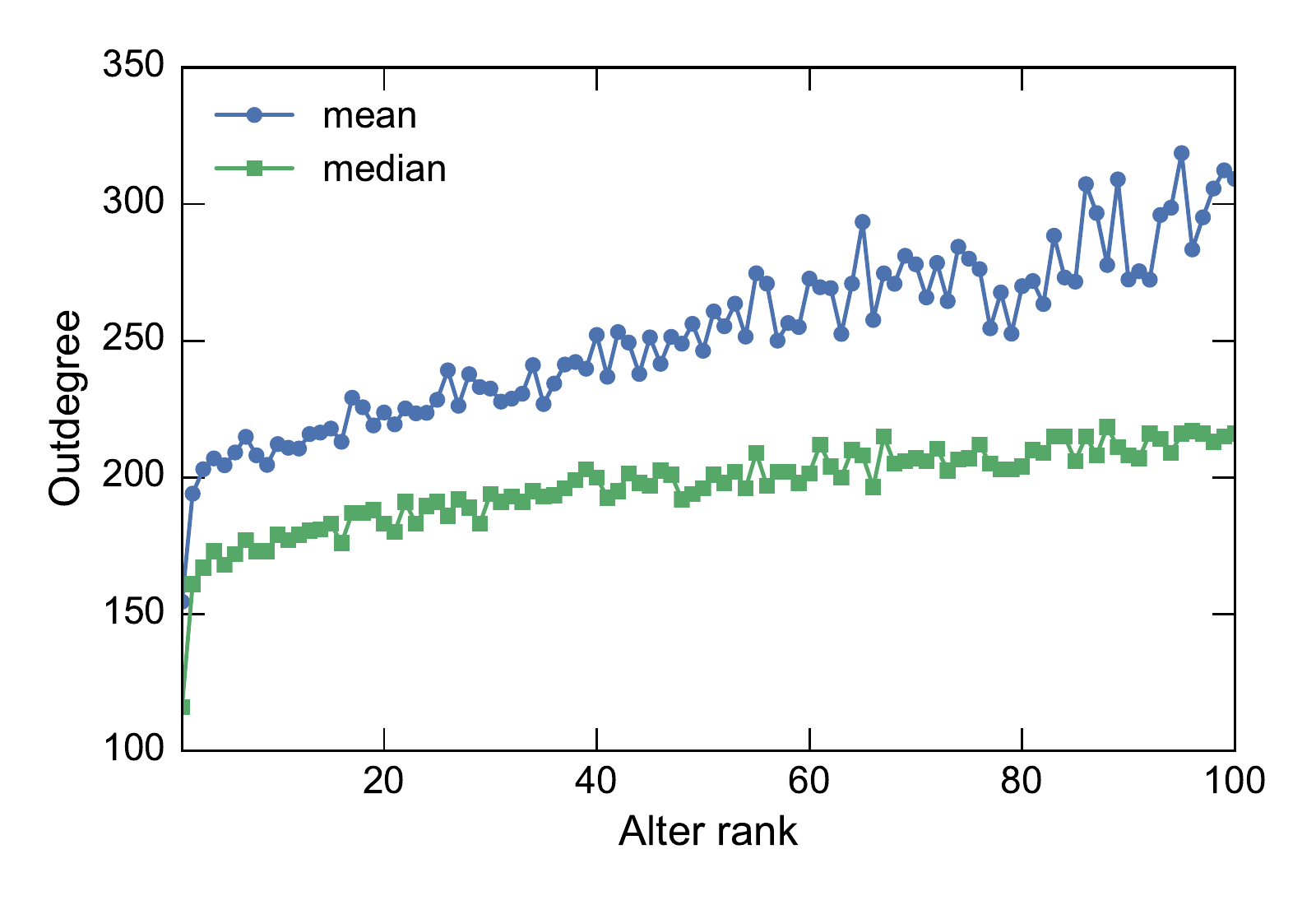}
        \caption{Mobile Phone\label{subfig:mainResultPhone}}
    \end{subfigure}
    \caption{Your friends are more popular than you, but your most frequently contacted alters are the least popular. Rank 1 = most frequently contacted.
        \label{fig:mainResult}}
\end{figure*}

The friendship paradox states that the mean degree of alters is greater than the degree of the ego for most egos in a social network. Using the data and our choices for outdegree, we found that the friendship paradox holds in both datasets. For Twitter it is highly prevalent, holding for 99.7\% of egos, comparable to that found by other researchers on Twitter datasets~\cite{Hodas13icwsm}. (It also holds for 99.3\% of egos when comparing the median outdegree of alters to the ego.)
The friendship paradox is prevalent in the mobile phone data as well, with  76.8\% of egos having lower outdegree than the mean outdegree of their alters. This percentage also matches closely with the 74\% identified for one of the datasets analyzed in the original friendship paradox work~\cite{feld1991your}.

We compared the outdegrees of alters as a function of how frequently the ego contacted them, as quantified by the rank of the alter, with rank-1 being the most frequently contacted alter. Ranking alters allows us to make comparisons across  egos of differing overall activity levels.  The mobile phone data allows us to examine a larger number of alters per ego (up to 100) whereas the Twitter data collection API effectively limited us to at most 15 alters.

Figure \ref{fig:mainResult} displays the relationship between alter $\kout$  and  alter rank. 
Despite their very different  natures, usage patterns, and social roles fulfilled by Twitter and mobile phone communication activity, as well as the different scopes of the data we have collected, we observed the same qualitative trend: the more frequently contacted alters tend to be of lower degree than the less frequently contacted alters. 
Indeed, in the mobile phone dataset,
the mean ego outdegree for the egos (166.3) is in fact slightly \emph{greater} than that for the rank-1 alter (154.2),
reversing the friendship paradox.
Only 48\% of egos have outdegree lower than their rank-1 alters,
in stark contrast to Feld's original finding that 74\% of egos have outdegree less than the mean outdegree for all of their alters.
Similarly, for the mobile phones the median ego outdegree is 120,
compared with 116 for the rank-1 alter.
While small, this difference is significant 
($p < 10^{-6}$, Wilcoxon signed-rank test), 
meaning in our mobile phone dataset we can reasonably claim
``your best friend is no more popular than you''.
Therefore, we have shown that, at least in these datasets, 
the main drivers of the friendship paradox are not the ``best friends'', but the more distant contacts.

\begin{figure*}[t!]
    \centering
    \begin{subfigure}[t]{0.5\textwidth}
        \centering
        \includegraphics[width=\textwidth]{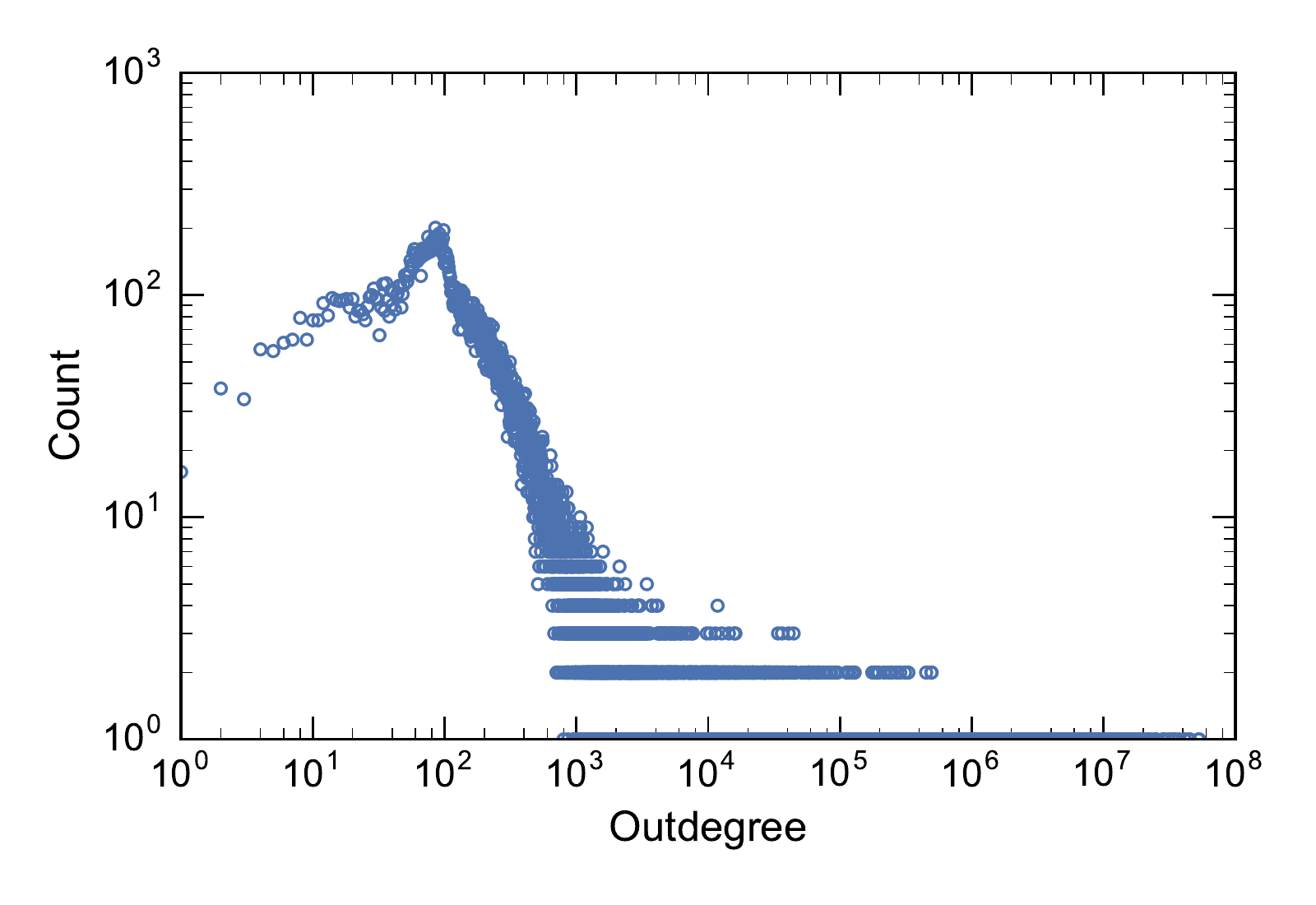}
        \caption{Twitter}
    \end{subfigure}%
    ~ 
    \begin{subfigure}[t]{0.5\textwidth}
        \centering
        \includegraphics[width=\textwidth]{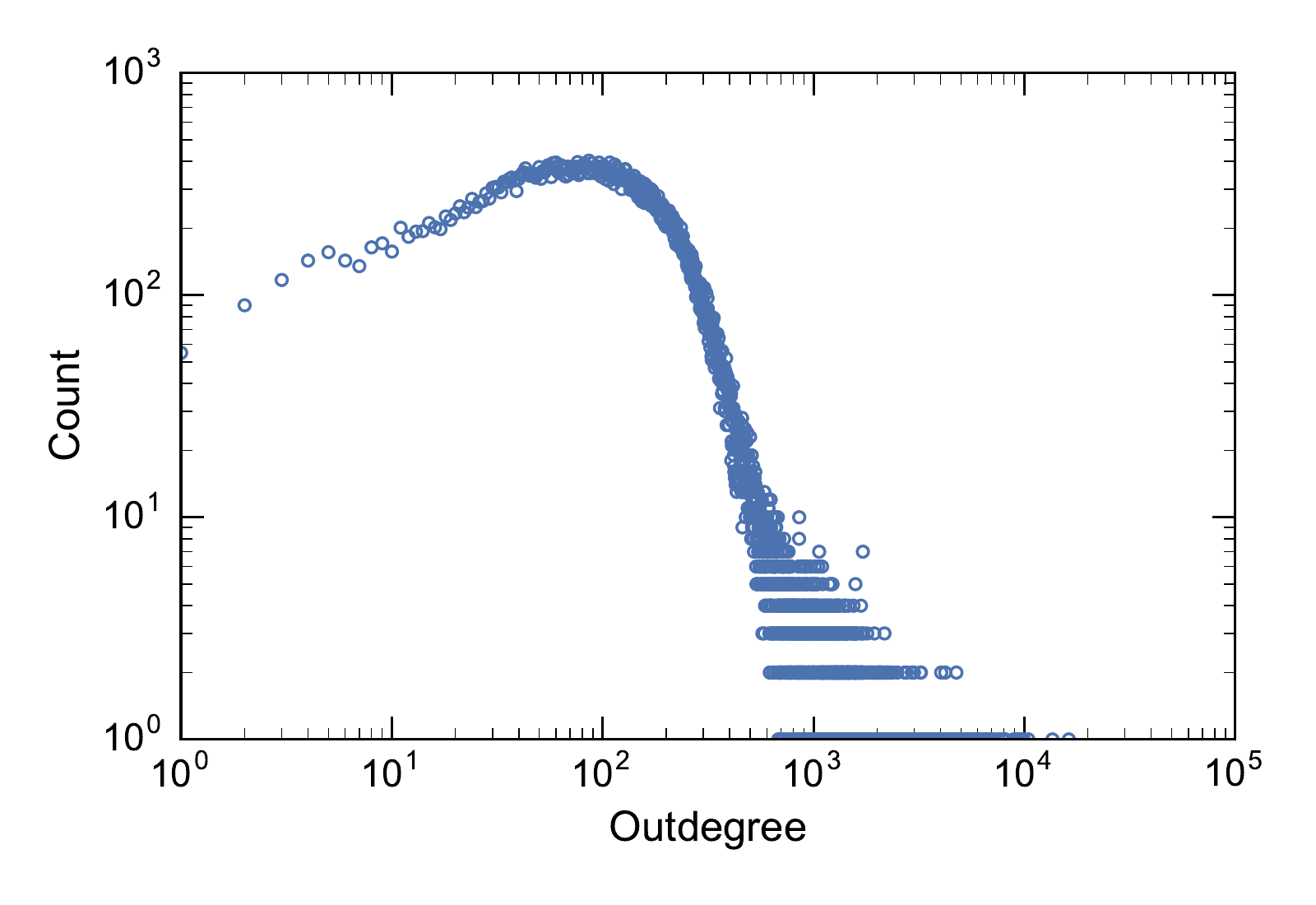}
        \caption{Mobile Phone\label{fig:degreeDistr_phone}}
    \end{subfigure}
    \caption{Degree distributions both peak near Dunbar's number of $k \approx 150$ while the Twitter degree distribution is more right-skewed than Mobile Phone.\label{fig:degreeDistrs}}
\end{figure*}

While the trend of increasing $\kout$ with increasing rank holds for both Twitter and Mobile Phone, there are differences. In particular, the extent of the increase is greater for Twitter: the median $\kout$ for rank-1 alters was an order of magnitude smaller than it was for rank-15 alters. In comparison, the change in $\kout$ is far more gradual for mobile phone egos. (Note that Fig.~\ref{subfig:mainResultTwitter} has a logarithmic vertical scale while Fig.~\ref{subfig:mainResultPhone} does not.)
The main driver behind this difference is the relative skewness of the degree distributions of the two social networks (Fig.~\ref{fig:degreeDistrs}). 
While the two distributions look similar in shape overall---and both peak at $\kout\approx100$, a value comparable with Dunbar's number of 150 \cite{dunbar1993coevolution}---the hubs on Twitter are far higher degree, nearly four orders of magnitude more than the mobile phones. 
This makes sense because of the ``broadcast'' nature of social media relative to a true social network: a mobile phone user simply cannot contact so many people, but online social media enables such large audiences. 
Because of this skew in $\kout$ we focused on the median outdegree in Fig.~\ref{subfig:mainResultTwitter}. 
Indeed, when considering the mean outdegree instead (Fig.~\ref{subfig:mainResultTwitter} inset), the trend is swamped by outliers. Both mean and median measures show similar results for mobile phones (Fig.~\ref{subfig:mainResultPhone}).

\begin{figure*}[t!]
    \centering
    \begin{subfigure}[t]{0.5\textwidth}
        \centering
        \includegraphics[width=\textwidth]{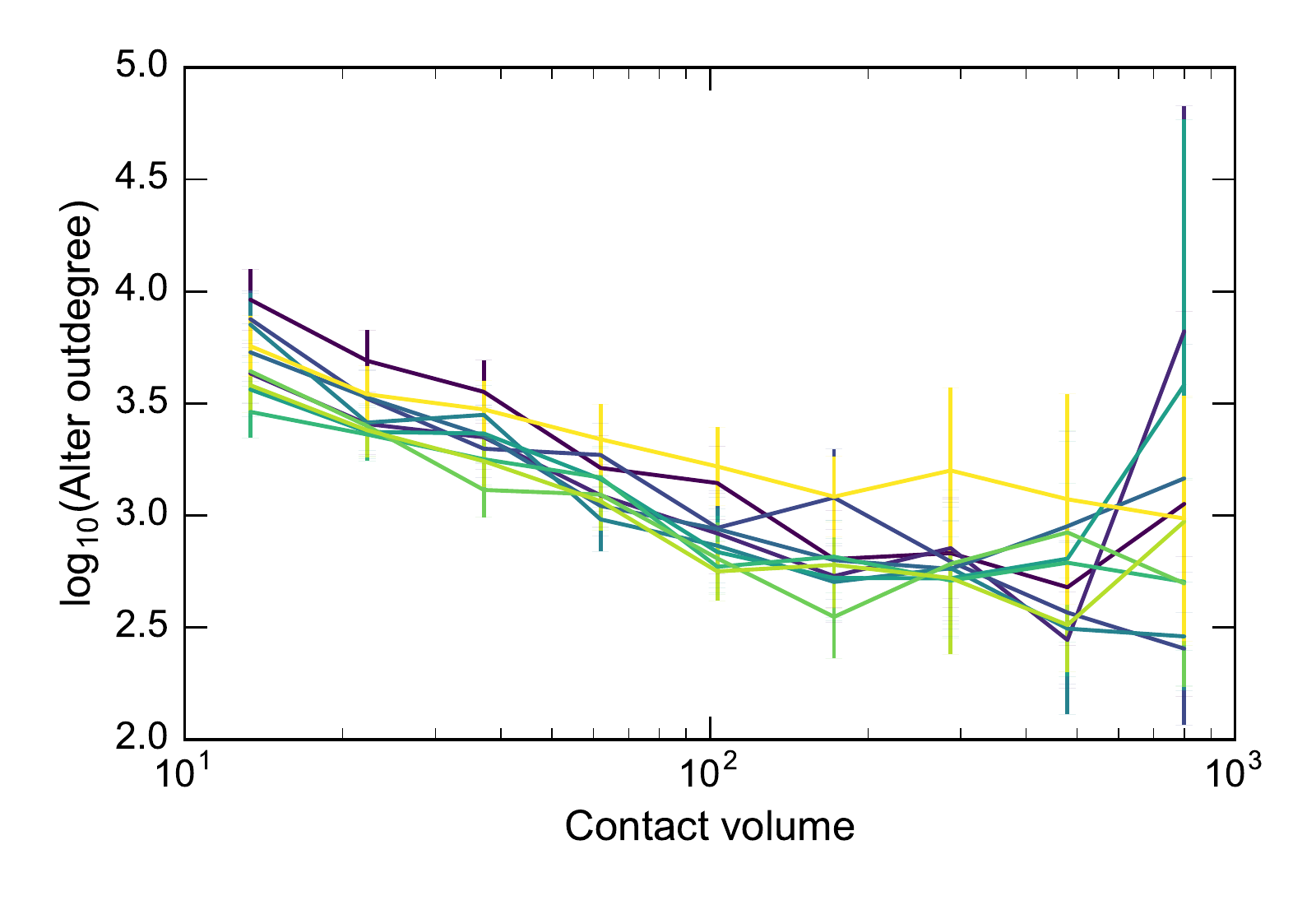}
        \caption{Twitter}
    \end{subfigure}%
    ~
    \begin{subfigure}[t]{0.5\textwidth}
        \centering
        \includegraphics[width=\textwidth]{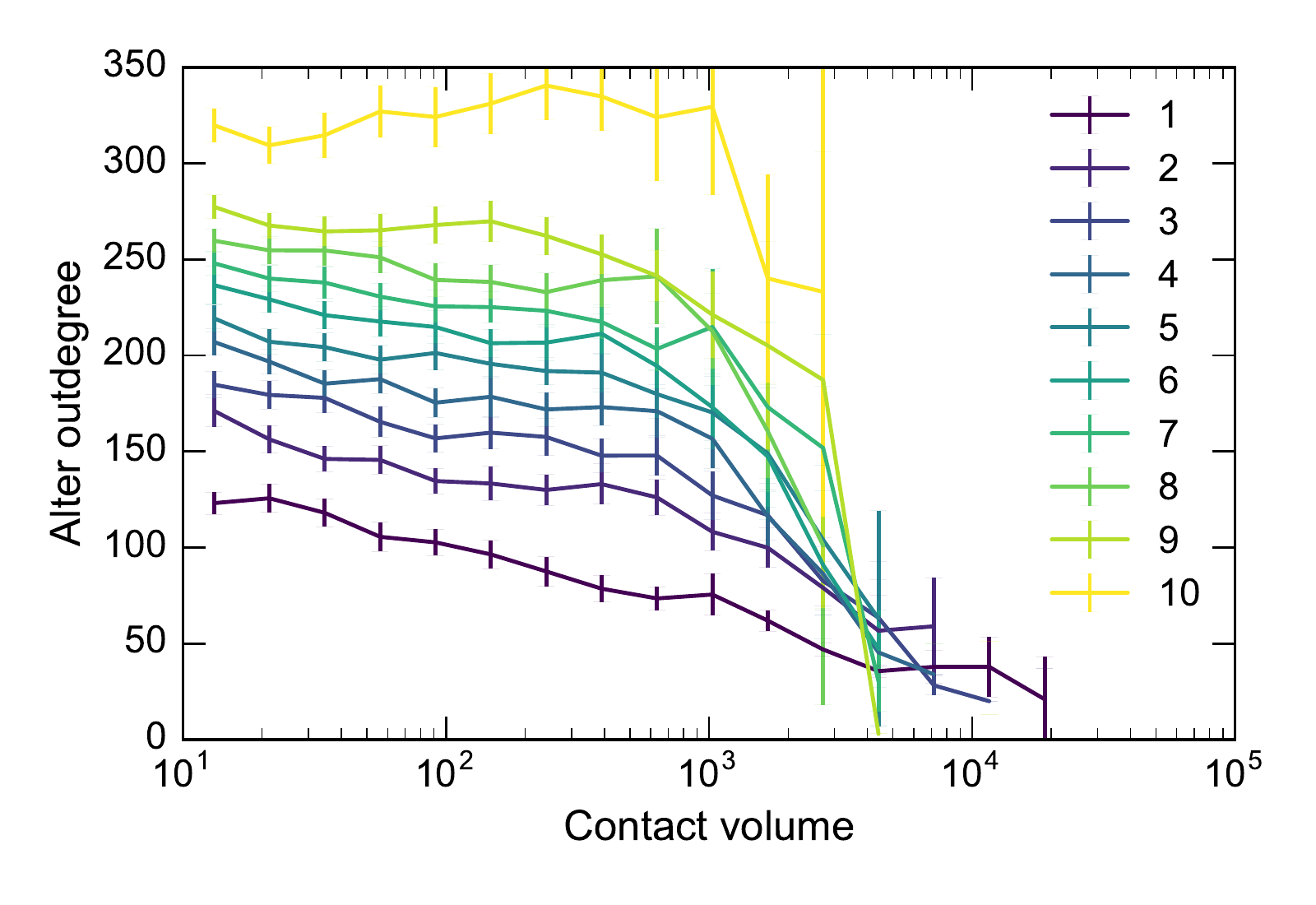}
        \caption{Mobile Phone}
    \end{subfigure}
    \caption{Direct (unranked) relationship between ego-to-alter contact volume and alter outdegree. Curves correspond to deciles of the ego outdegree distribution (colors the same in both panels).
    \label{fig:directRelationshipContactVolumeOutdegree}}
\end{figure*}

Figure \ref{fig:directRelationshipContactVolumeOutdegree} shows the relationship between decreasing contact volume and increasing alter outdegree directly, without ranking alters.
Ranking is a useful technique here because it eliminates much of the \textit{heterogeneity} across egos: very active egos and less heavily active egos will have very different overall contact volume, but will appear similar in ``rank space''. 
Despite this, it is important to also look at the direct relationship between contact volume and alter outdegree. We took one step in Fig.~\ref{fig:directRelationshipContactVolumeOutdegree} to account for heterogeneity---we have binned the data by deciles of the outdegree degree of egos (which strongly correlates with the total contact volume of the ego). 
Due to the skew in outdegree of alters in Twitter (Fig.~\ref{fig:degreeDistrs}) we first log-transformed the outdegree for Twitter.
We see for both datasets a significant decreasing trend. With the possible exception of the 10th decile of mobile phone egos, increases in contact volume overwhelmingly correlate with decreases in alter outdegree.

\begin{figure*}
    \centering
    \begin{subfigure}[t]{0.5\textwidth}
        \centering
        \includegraphics[width=\textwidth]{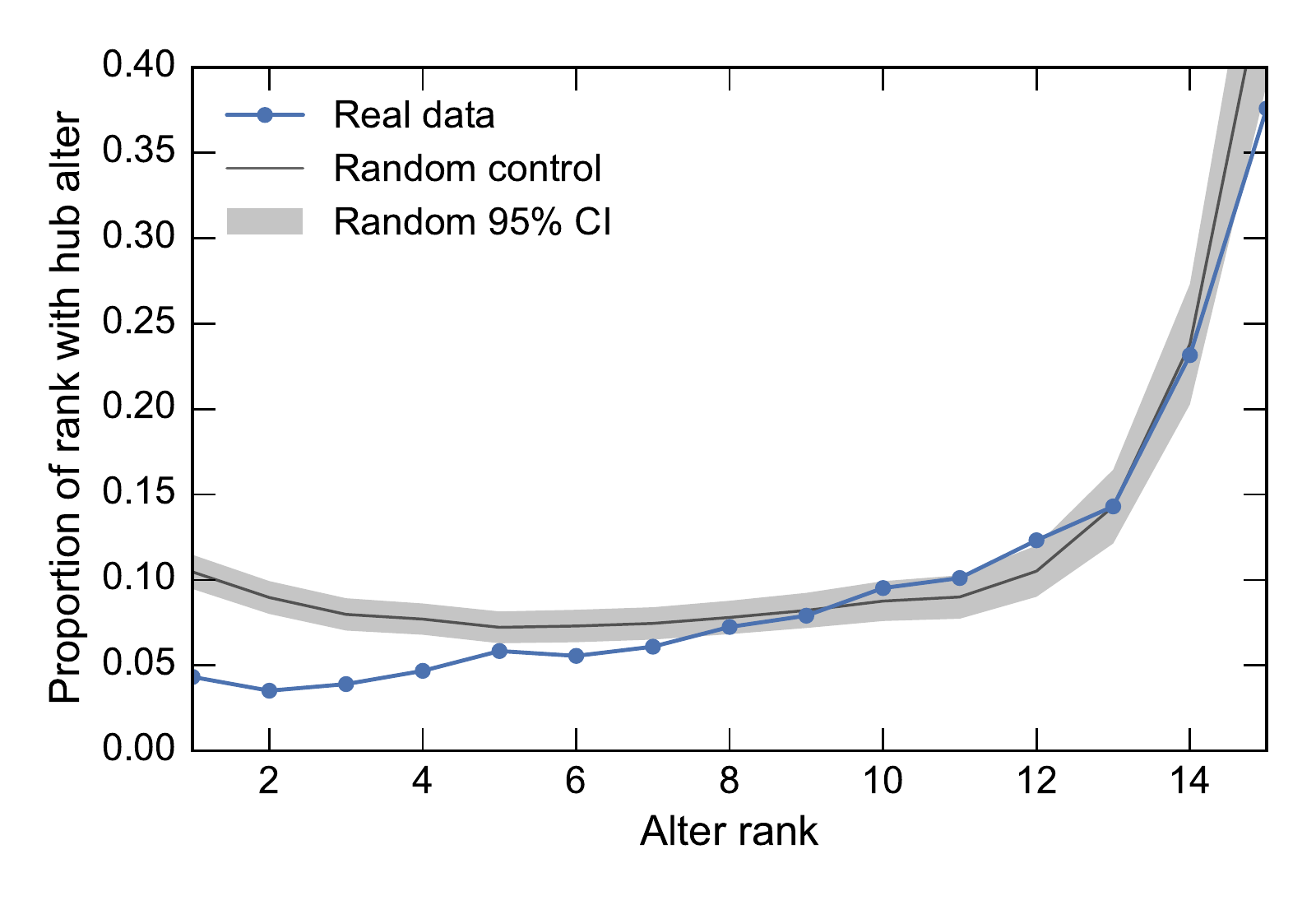}
        \caption{Twitter}
    \end{subfigure}%
    ~ 
    \begin{subfigure}[t]{0.5\textwidth}
        \centering
        \includegraphics[width=\textwidth]{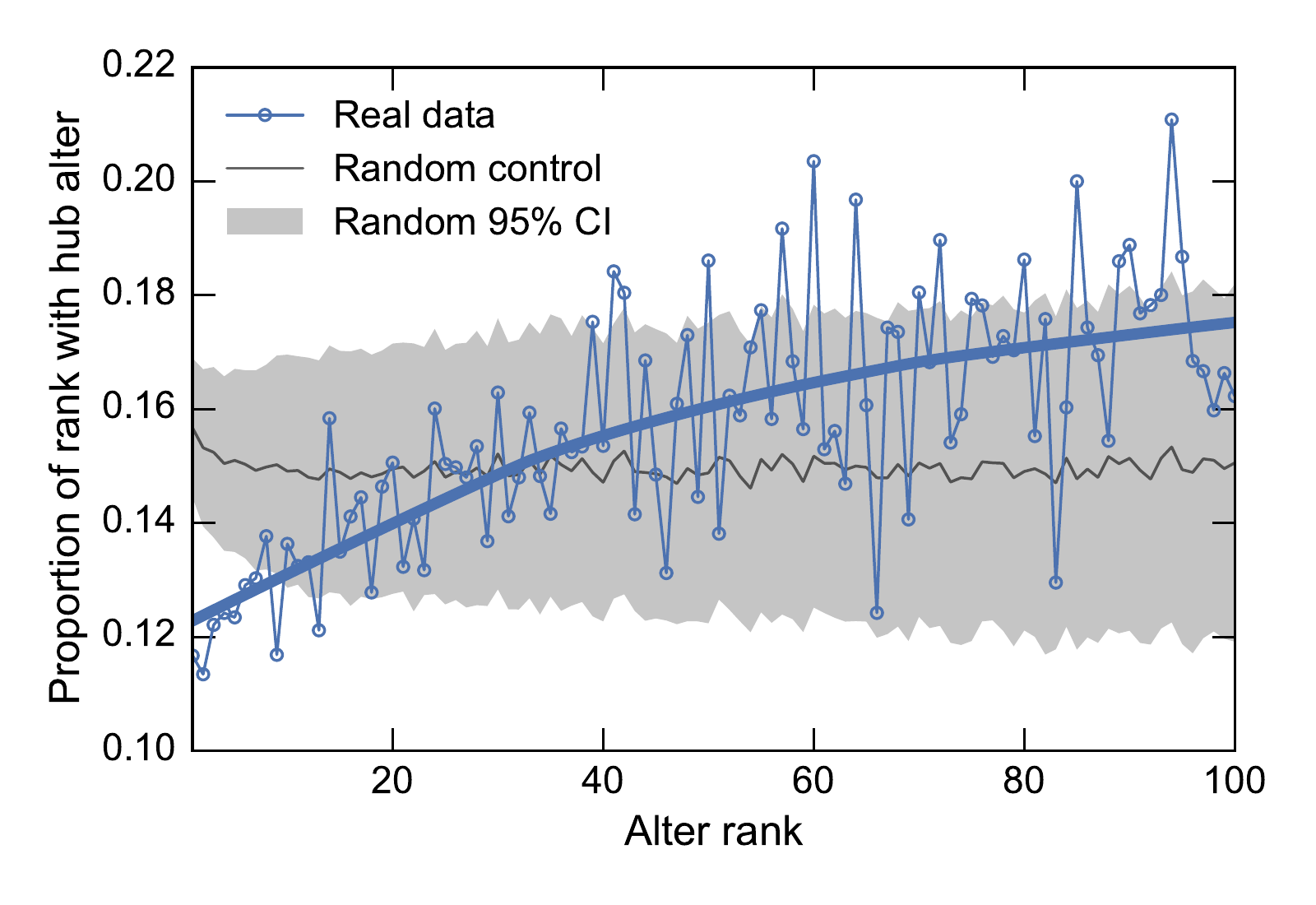}
        \caption{Mobile Phone}
    \end{subfigure}
    \caption{The proportion of ego-alter dyads where the alter is the ego's ``hub alter''. In both datasets, the likelihood
    of a hub alter increases with rank and the likelihood for a hub alter in the first $\approx 7$ ranks is significantly lower than expected by chance. The guide line in (b) is a LOESS curve.
    \label{fig:dyadHubAltersProportion}
    }
\end{figure*}

To further investigate the link between contact volume and alter outdegree, in particular to move away from some of the statistical averaging (mean or median $\kout$) applied so far, we defined the \textbf{hub alter} of an ego as the alter with the maximum $\kout$ (excluding any alters whose outdegree was unavailable in the data). 
We investigated where the hub alter tends to fall. Is the hub alter more likely to occur at lower ranks, among the heavily contacted alters? Or at higher ranks, among the less contacted alters.
To find out, we computed the proportion of ego-alter dyads for each rank $r$ where the alter at rank $r$ was the hub alter, as a function of $r$ (Fig.~\ref{fig:dyadHubAltersProportion})~\footnote{For this calculation we only considered egos with at least 5 alters with an available $\kout$. 
Egos with few alters must necessarily have the hub alter at lower ranks, and this filter criterion prevents this dataset limitation from masking any trends.}. 
In both datasets there is a significant, increasing trend (Twitter: Spearman's $\rho = 0.8321$, $p< 0.001$; Mobile Phone: $\rho = 0.7208$, $p<10^{-16}$), meaning that it is the less frequently contacted alters who tend to be the hub alter.

\textit{Permutation test---}
Is there a significant relationship between the hub alter and its rank (Fig.~\ref{fig:dyadHubAltersProportion}), or could it be explained by a random effect or bias in the data?
To test the significance of what rank the hub alters fall at, we need an appropriate null model. 
We assume that, if there is no relationship, each possible rank is equally likely to be the hub alter. 
This is a Monte Carlo permutation model that is easy to compute (we did 1000 independent permutations for each dataset). 
However, there is an important caveat: some alters were excluded from our data either due to insufficient written text (Twitter) or those alters are not subscribers to the service provider (Mobile Phone); see Datasets for details.
This means that we did not have their outdegrees even though for both data we ranked all the alters correctly. 
However, this issue is easy to bypass---since we did not use those alters we simply do not allow their ranks to be chosen randomly under the null. 
In other words, for an ego with $n$ total alters, the null model samples a rank at random not from all ranks $1, \ldots, n$ but only from the $0 < n_\mathrm{available} \leq n$ available ranks for that ego. 
Crucially, accounting for ``missing'' alters makes sure that any results we observe cannot be explained by sampling effects. 
We simulated these null models and overlaid the results on Fig.~\ref{fig:dyadHubAltersProportion} (shaded regions denote the middle 95\% of random realizations not confidence intervals on the mean proportion). 
For both datasets, there was almost no relationship between alter rank and the proportion of hub alters under the null, except for high ranks in the Twitter data (which is explainable by sampling effects discussed in Datasets). 
In fact, otherwise, if anything, there was actually a weak decreasing trend in the null model for the first few (5--10) alters of both datasets, likely due to averaging over egos with different numbers of alters. 
In contrast, the real data displays a strong increasing trend. 
Further, for both datasets, the proportion of hub alters for the first few  (approximately 7) ranks is significantly lower than expected from chance.

\begin{figure*}
    \centering
    \begin{subfigure}[t]{0.5\textwidth}
        \centering
        \includegraphics[width=\textwidth]{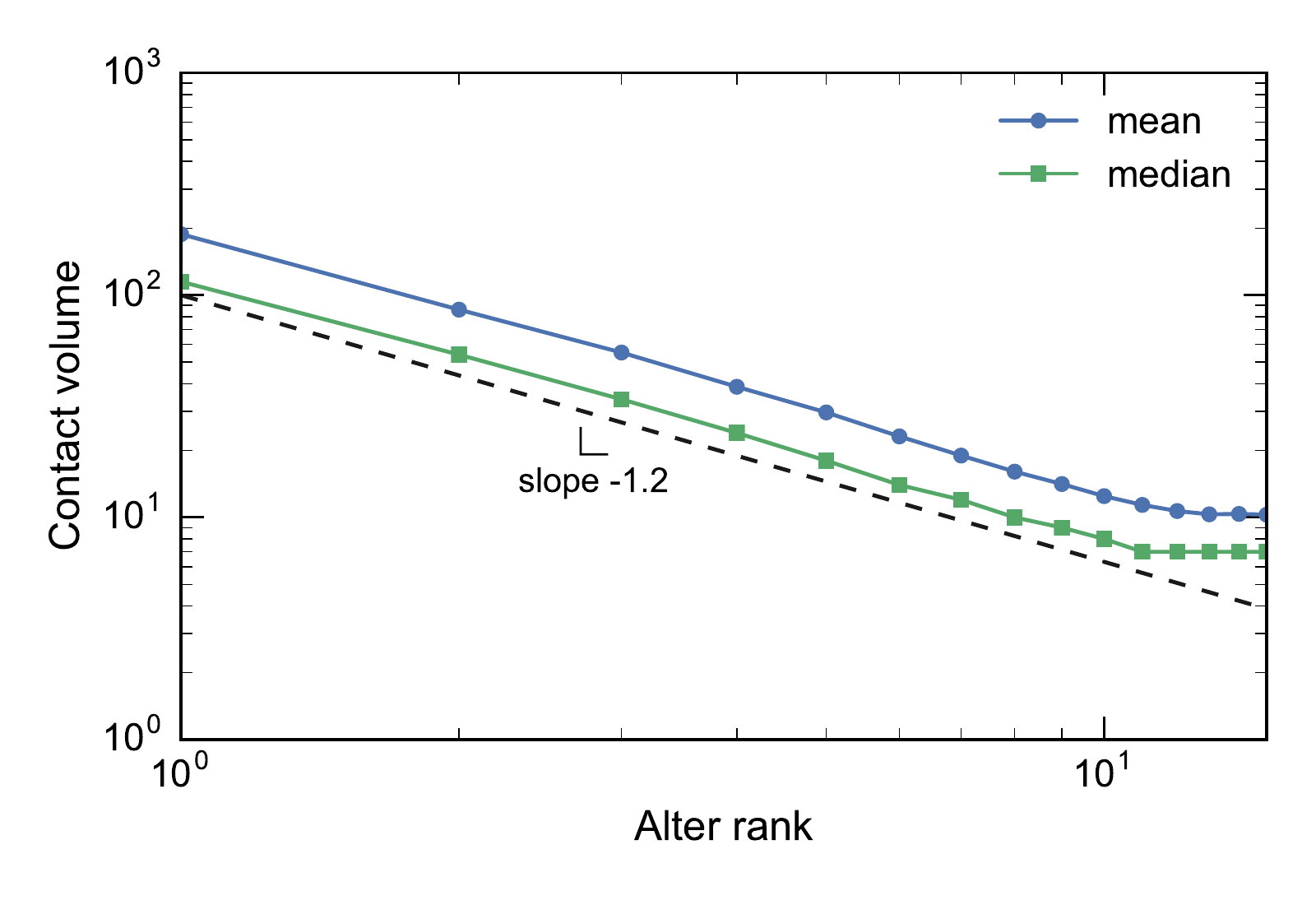}
        \caption{Twitter}
    \end{subfigure}%
    ~ 
    \begin{subfigure}[t]{0.5\textwidth}
        \centering
        \includegraphics[width=\textwidth]{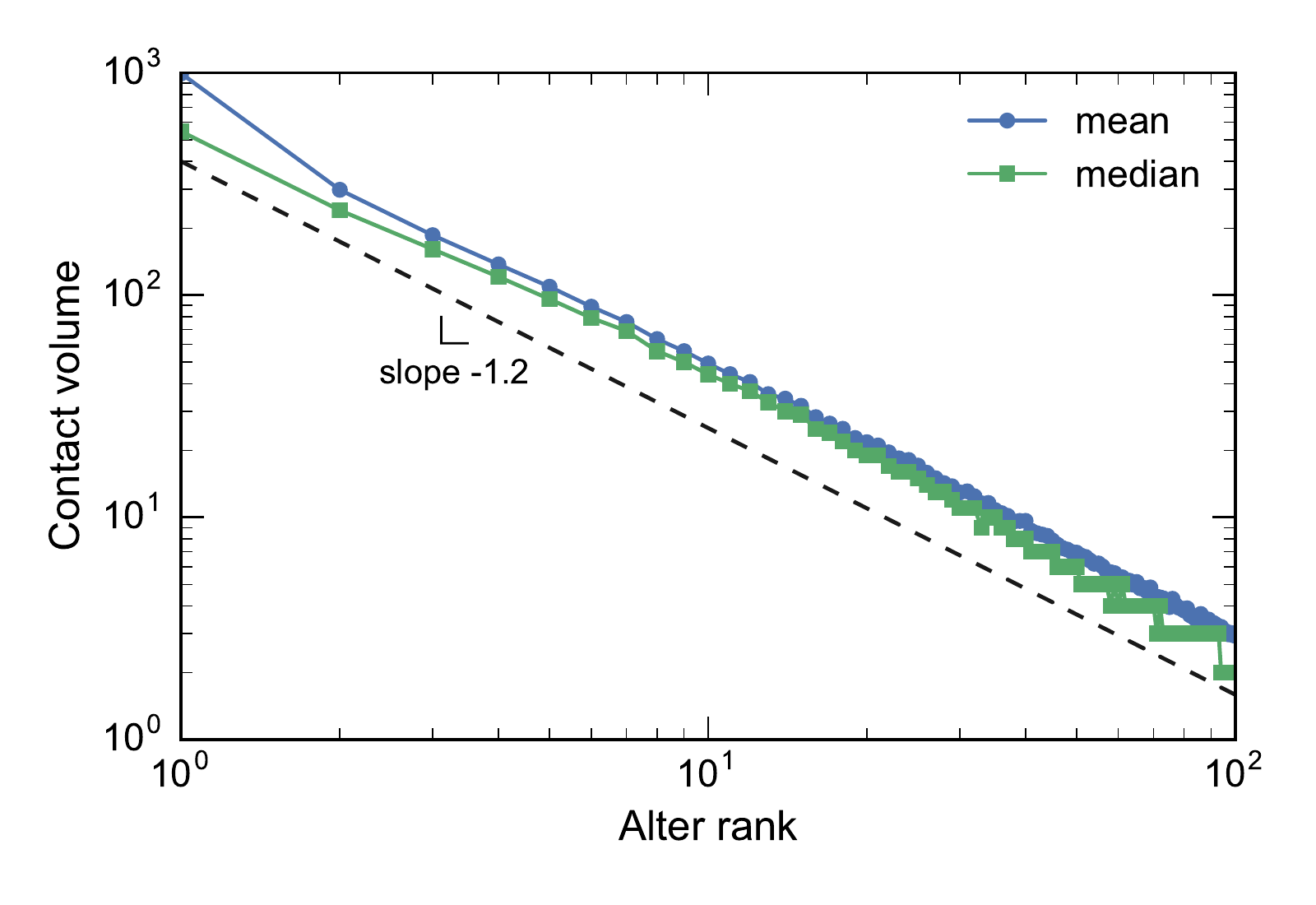}
        \caption{Mobile Phone}
    \end{subfigure}
 \caption{Contact volume (number of at-mentions, number of calls/text messages) versus alter rank. Note that the two plots have different aspect ratios and that the dashed guideline has the same slope for both: Contact volume = $C \times$ (Alter rank)$^{-1.2}$.
    \label{fig:contactVolume_alterRank}}
\end{figure*}

We also observed a Zipfian ranking relationship~\cite{zipf1949human} between contact volume and alter rank. In Fig.~\ref{fig:contactVolume_alterRank} we show that both datasets obey a power law relationship with power law exponent $\approx 1.2$ (although our Twitter data only provides a little over a decade of data).
This scaling has previously been observed for mobile phones when ranking locations by the number of visits~\cite{gonzalez2008understanding} and when ranking social ties by contact volume~\cite{bagrow2012mesoscopic} as we did here.
A similar scaling of $\approx 1.3$ was also found by Frank et al. for the probability of tweeting from a location as a function of location rank~\cite{Frank2013}.
Remarkably, despite their different natures, potential differences in how individuals use the respective communication services, and sampling differences in our collected data, the Twitter and Mobile Phone datasets demonstrate nearly identical scaling.

\begin{figure}
    \centering
    \begin{subfigure}[t]{0.5\textwidth}
        \centering
        \includegraphics[width=\textwidth]{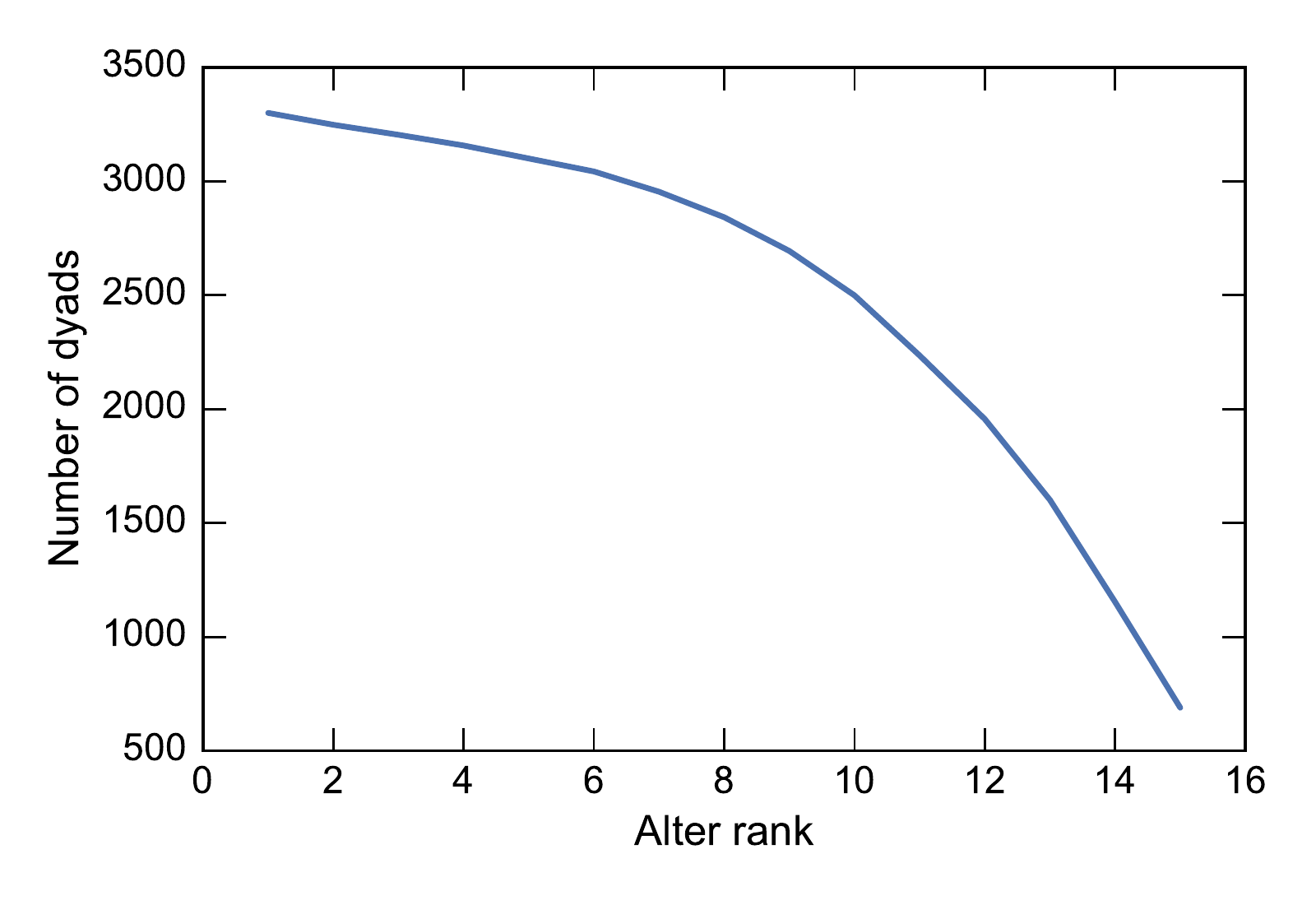}
        \caption{Twitter}
    \end{subfigure}%
    \\ 
    \begin{subfigure}[t]{0.5\textwidth}
        \centering
        \includegraphics[width=\textwidth]{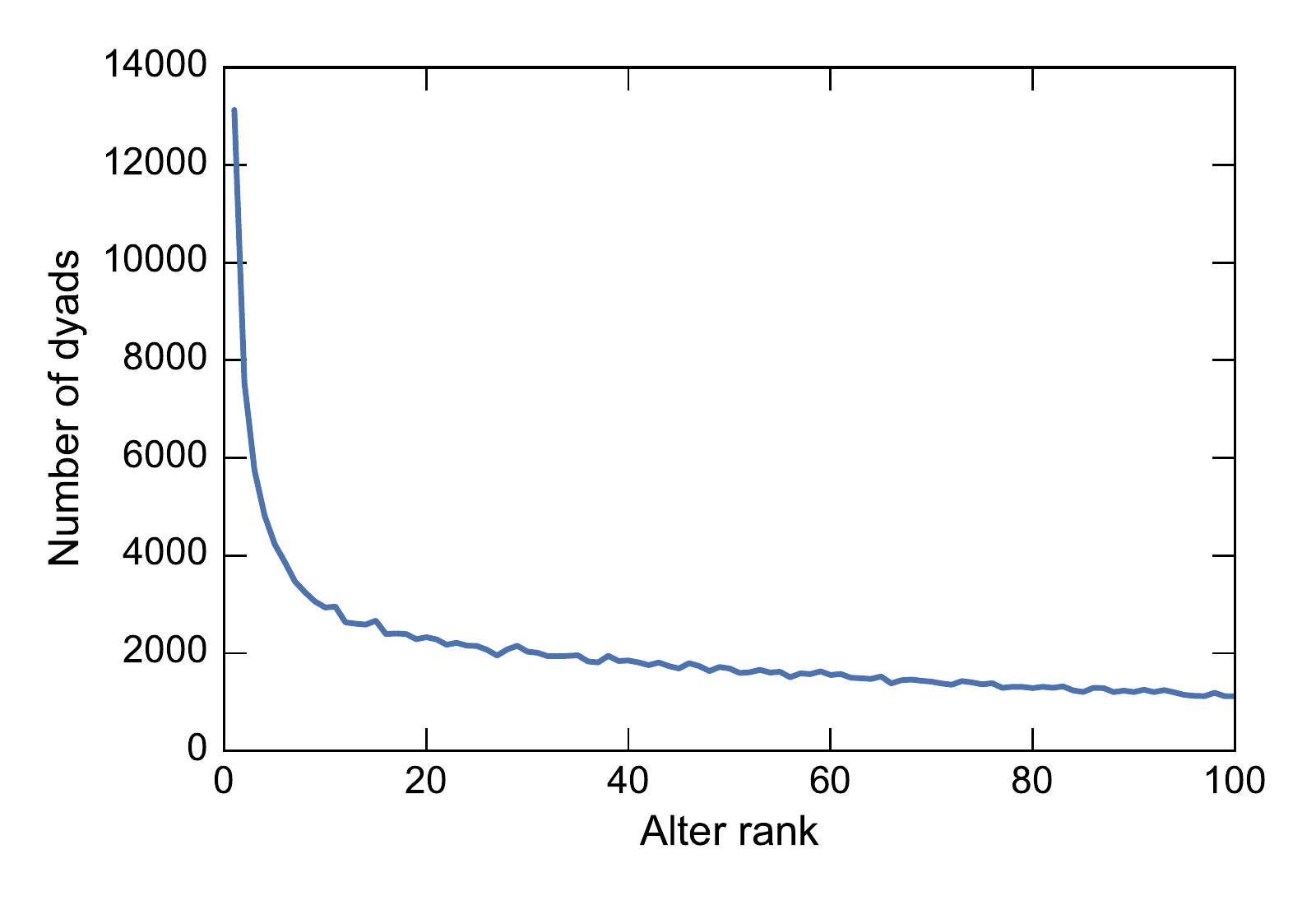}
        \caption{Mobile Phone}
    \end{subfigure}
    \caption{Number of ego-alter dyads across alter ranks.\label{fig:numberOfDyads}}
\end{figure}

A key difference between the Twitter and Mobile Phone data is sampling heterogeneity across ego-alter dyads. 
To quantify this difference,
we measured how many ego-alter pairs (dyads) were present in the data for each alter rank. This gives us the distribution of number of alters across egos (which may differ from ego $\kout$; see Datasets).
As shown in Fig.~\ref{fig:numberOfDyads}, most Twitter egos have comparable numbers of available alters, at least up to around 8 alters. In contrast, the Mobile Phone egos, while potentially having up to 100 alters, tend to have fewer than ten alters and unlike Twitter this number drops off quickly. 
There are still at minimum hundreds (Twitter) or thousands (Mobile Phone) of dyads available at every alter rank.
This points to distinct differences in the form of the social network on these two communication platforms and how the data is collected. Yet, despite these differences, we have already observed a number of similarities between Twitter and Mobile Phone in how egos relate to alters.

%

\subsection{Information cascades}

The friendship paradox, in concert with the strength of weak ties and Dunbar's number, plays a significant role in information diffusion on social networks. 
To explore how  the contact volume--outdegree relationship affects information transfer,
we used a simple dynamical model of information transfer on the mobile phone network.
We generated a network with 88,137 nodes and 8,774,126 edges from the degree distribution shown in Fig.~\ref{fig:degreeDistr_phone} using the configuration model~\cite{molloy1995critical,molloy1998size} as implemented in the Python NetworkX package \cite[ver.\ 1.11]{schult2008exploring}.
Starting from an initial infected node, 
we then ran outbreaks using a basic Susceptible-Infected model~\cite{kermack1927contribution,anderson1992infectious,Newman2010} under two different transmission regimes:
\begin{enumerate}[(i)]
\item \emph{uniform transmission without contact volume dependence}: 
transmission probability $\beta = 0.01$ is held constant across all nodes;
\item \emph{alter-dependent transmission with contact volume dependence}: 
each ego $i$ has a per-alter transmission probability $\beta_i = C_i/r$ inversely proportional to alter rank $r$, similarly to Fig.~\ref{fig:contactVolume_alterRank}.
\end{enumerate}
To set equal the expected number of secondary infections from an infected node across experiments, we choose
\begin{equation}
C_i = \frac{n_i}{H(n_i)}\beta
\end{equation}
where $H(n_i)$ is the harmonic number and $n_i$ is the number of alters of ego $i$.
Furthermore, inspired by the trends shown in Fig.~\ref{fig:dyadHubAltersProportion}, we defined alter rank in the model network by the outdegree of the alter, so that the highest-ranked alters were the least ``famous''. 
Recognizing that this represents a particularly strong version of the contact volume--alter rank relationship where high-ranked alters are \emph{always} more famous,
we also introduced a parameter $p$ which controls the strength of this effect:
For each potential infection during an outbreak, we infected a new node as per scenario (i) with probability $1-p$ and otherwise as per scenario (ii) with probability $p$.
This has the effect of adding some randomness into the contact volume-alter relationship, 
making the ranking less strictly enforced in the numerical experiments.
Overall, scenario (i) provides a baseline and comparative control, 
while scenario (ii) captures rough approximations of the dynamical ingredients we have observed here.

\begin{figure*}
    \centering
    \begin{subfigure}[t]{0.5\textwidth}
        \centering
        \includegraphics[width=\textwidth]{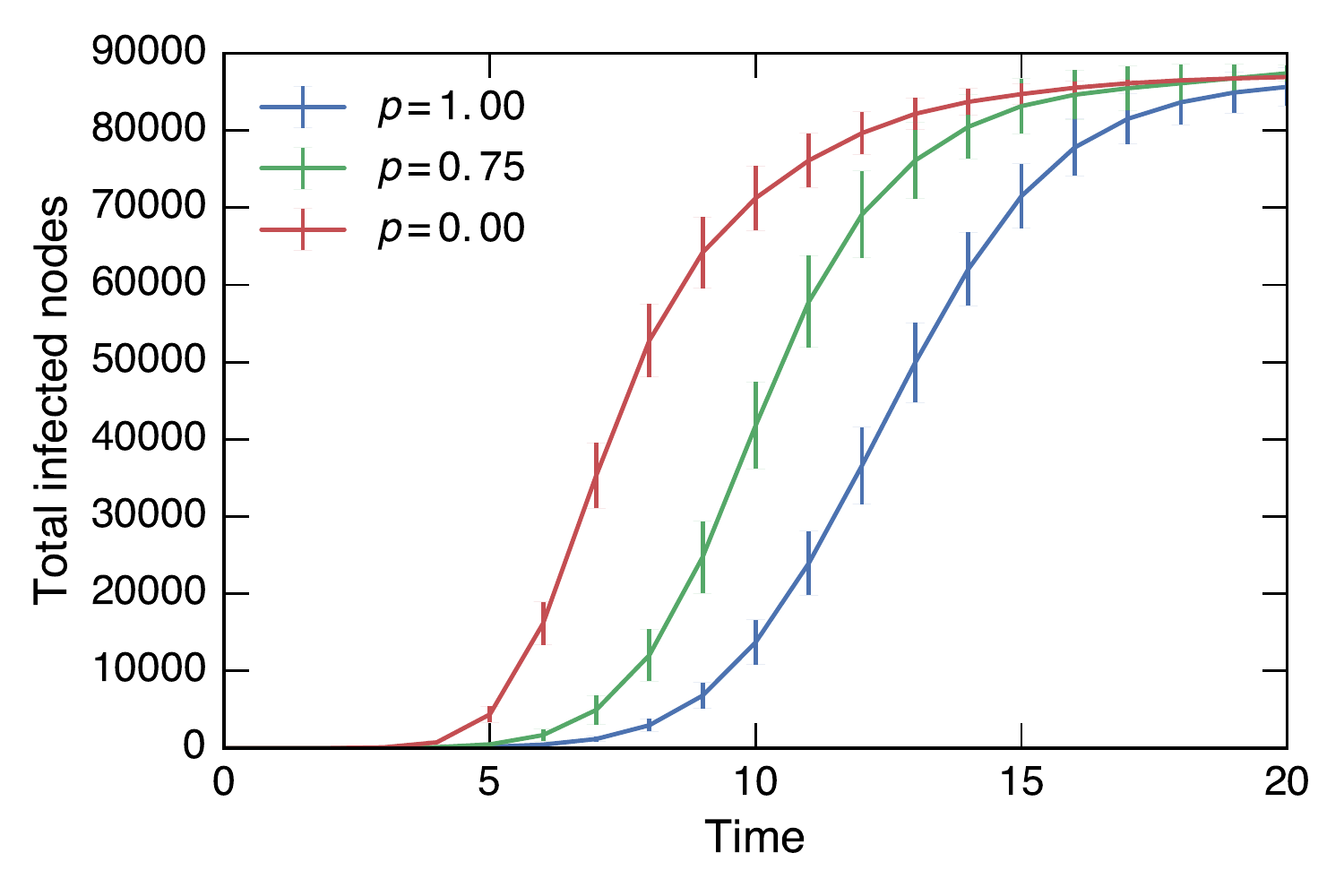}
        \caption{Total infections}
    \end{subfigure}%
    \begin{subfigure}[t]{0.5\textwidth}
        \centering
        \includegraphics[width=\textwidth]{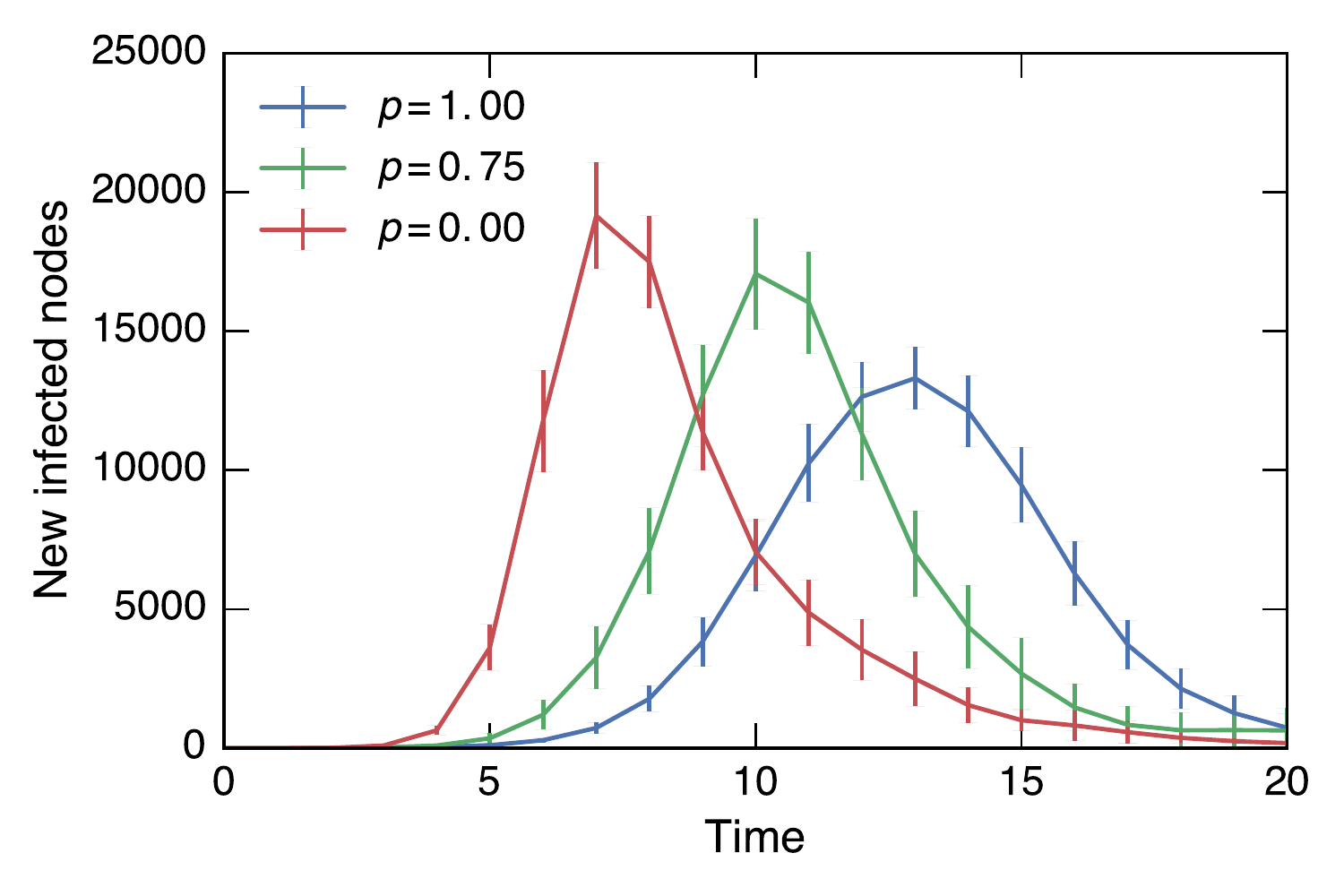}
        \caption{New infections}
    \end{subfigure}
    \caption{Epidemic curves for friendship paradox with (blue) and without (green) contact volume dependence.
\label{fig:epidemic_curves}}
\end{figure*}

We ran 100 independent simulations of 20 time steps of an outbreak using synchronous updating on the same network, starting from a single infected node.
Figure \ref{fig:epidemic_curves} shows the results without contact volume $p = 0$ (red), with contact volume relationship $p = 1$ (blue), and for an intermediate value $p = 0.75$ (green).
Error bars show 95\% confidence intervals on the simulation averages.
Despite the two different models having the same expected number of secondary infections from any infected node, the contact volume dependence dramatically reduces the rate of information propagation over the network, as high-degree nodes are much less likely to become infected by a low-degree node.

\section{RELATED WORK}


The key observation of the friendship paradox is that, because highly popular individuals are over-represented across the sets of alters of egos, the average degree of alters is right-skewed, leading to the colloquial ``your friends have more friends than you''~\cite{feld1991your}. 
Perceptions can be biased because of the paradox~\cite{zuckerman2001makes} and and personality traits which correlate with network degree, such as extroversion~\cite{pollet2011extraverts} will be influenced by the paradox.

The original observations of the friendship paradox have led to a number of papers investigating it and similar paradoxes and how they affect the views of individuals within a social network.
The original author showed that the variation in classroom sizes exhibits \cite{feld1977variation}, while Ugander et al.~showed that the friendship paradox exists at a very large scale through analyzing the anatomy of the Facebook social graph \cite{ugander2011anatomy}.
Generalizations of the friendship paradoxes have also been studied~\cite{eomGeneralizedFriendshipParadox2014}. That work shows that any observable quantity which correlates with degree will be biased in the same way as degree itself.
Interestingly, the friendship paradox has also been proposed as an exploitable feature of networks.
Christakis and Fowler used the friendship paradox to suggest a social monitoring system for early detection of influenza outbreaks \cite{Christakis2010}. They also investigated this in Twitter data for information propagation~\cite{GarciaHerranz2014}.
Dynamics of communication have been studied via telecommunications data by many researchers. For example,
Karsai et al.\ and Borge-Holtheofere et al.\ studied bursty communication dynamics and information transfer in mobile phone and other datasets~\cite{karsai2011small,mocanu2015collective,BorgeHolthoefere2016}, including egocentric networks~\cite{karsaiCorrelatedEgo2012} but did not investigate the friendship paradox.
To the best of our knowledge, research has not jointly investigated the relationships between contact volume or other individuals' contact behavior with the friendship paradox and its effects.

The work which is perhaps most similar to our own is the work of Hodas et al.~\cite{Hodas13icwsm}. They showed that, not only does the paradox hold in Twitter, it is very strong, affecting $\approx 99\%$ of egos, as we also found. In addition, they investigated several other related paradoxes, including that alters are more active (activity paradox) and that alters are exposed to more viral tweet content (virality paradox). This shows that both the structure of the Twitter follower graph and the dynamics of viral content are influenced by paradoxes such as the friendship paradox.
Their work was limited to Twitter only and considered numbers of tweets and retweets, but unlike the work here did not investigate contact volume (at-mentions) between users.

Most work on the friendship paradox has been structural in nature, comparing the static degrees of egos and alters. Yet, social networks host dynamic activity patterns, and egos will maintain very different levels of contact among their social ties, due to differing degrees of social intimacy, cognitive limits on the size of a social network, and simply because of time constraints: there are only so many hours in the day. The work we present here moves beyond the static network and strengthens the connections between the friendship paradox, the strength of weak ties, and Dunbar's number.

\section{CONCLUSIONS}

We have investigated how the strength of the friendship paradox relates to contact volume between individuals in two real, social network datasets.
Specifically, we have found that while ``your friends are more famous than you'' holds on average in the overwhelming majority of cases,
this effect is in general far less pronounced for your ``closest friends'' or most frequently contacted social ties.
Instead, more distant friends are more often the main drivers of the friendship paradox.
In general, more frequently contacted alters are of lower outdegree,
and conversely an ego's ``hub alter'', the highest outdegree alter, is more likely to be less frequently contacted by the ego.
Using a conceptual Susceptible-Infected (SI) model of information transmission we demonstrated numerically that the contact volume-alter rank inverse relationship can reduce the rate of information transfer over the network.
Future work will look in more detail at the content of the messages passing between egos and alters with a view towards characterizing social information flow.

The contact volume-outdegree relationship may be relevant in epidemiology, 
where ``acquaintance immunization'' \cite{Cohen2010} has been shown to be an effective strategy. 
If individuals actually end up immunizing their closest, or high-contact volume alters, 
then this immunization strategy may in fact turn out to be less effective than if they immunized more distant friends.

Online communication provides the means to vastly increase the apparent size of a  user's social circle, leading to information overload from the social inputs of all the newly available alters. 
Yet, if egos preferentially contact the lowest degree alters as we have studied, 
then this information overload may be attenuated. 
Further investigation of information transfer under this relationship is warranted, in particular how it may relate to the formation of filter bubbles or echo chambers.






\section*{ACKNOWLEDGMENT}

JPB and CMD acknowledge support by the National Science Foundation under Grant No.\ IIS-1447634.
LM acknowledges support from the Data To Decisions Cooperative Research Centre (D2D CRC),
as well as from the 
ARC Centre of Excellence for Mathematical and Statistical Frontiers (ACEMS).



\bibliographystyle{IEEEtran}
\bibliography{references}

\end{document}